\documentclass[aps,prl,reprint,amsmath,showpacs,superscriptaddress]{revtex4-1}
\usepackage{graphicx}
\usepackage{textcomp}
\usepackage{float}

\usepackage{xcolor}
\usepackage[normalem]{ulem}
\usepackage{upgreek}

\newcommand{\vF}{v_{\mathrm{F}}}        % vF

\newcommand{\wG}{\omega_\mathrm{G}}
\newcommand{\wD}{\omega_\mathrm{2D}}
\newcommand{\WG}{\Gamma_\mathrm{G}}

\newcommand{\EF}{E_\mathrm{F}}
\newcommand{\mum}{\mathrm{\upmu m}}
\newcommand{\nm}{\mathrm{nm}}
\newcommand{\cm}{\mathrm{cm^{-1}}}
\newcommand{\etal}{\textit{et~al.~}}     % Italicized "et al."

\begin{document}

\title{Electrical Control over Phonon Polarization in Strained Graphene
}

\author{J.~Sonntag}
	\email{Corresponding author: sonntag@physik.rwth-aachen.de}
    \affiliation{JARA-FIT and 2nd Institute of Physics, RWTH Aachen University, 52074 Aachen, Germany}
    \affiliation{Peter Gr{\"u}nberg Institute (PGI-9), Forschungszentrum J{\"u}lich, 52425 J{\"u}lich, Germany}

\author{S. Reichardt}
    \affiliation{Department of Physics and Materials Science, University of Luxembourg, L-1511 Luxembourg, Luxembourg}

\author{B. Beschoten}
    \affiliation{JARA-FIT and 2nd Institute of Physics, RWTH Aachen University, 52074 Aachen, Germany}

\author{C.~Stampfer}
    \affiliation{JARA-FIT and 2nd Institute of Physics, RWTH Aachen University, 52074 Aachen, Germany}
    \affiliation{Peter Gr{\"u}nberg Institute (PGI-9), Forschungszentrum J{\"u}lich, 52425 J{\"u}lich, Germany}

\date{\today}

\begin{abstract}
We explore the tunability of the phonon polarization in suspended uniaxially strained graphene by magneto-phonon resonances.
The uniaxial strain lifts the degeneracy of the LO and TO phonons, yielding two cross-linearly polarized phonon modes and a splitting of the Raman G~peak.
We utilize the strong electron-phonon coupling in graphene and the off-resonant coupling to a magneto-phonon resonance to induce a gate-tunable circular phonon dichroism.
This, together with the strain-induced splitting of the G~peak, allows us to controllably tune the two linearly polarized G~mode phonons into circular phonon modes.
We are able to achieve a circular phonon polarization of up to 40~\% purely by electrostatic fields and can reverse its sign by tuning from electron to hole doping.
This provides unprecedented electrostatic control over the angular momentum of phonons, which paves the way toward phononic applications.
\end{abstract}

\maketitle

Phonons -- collective excitations of lattice vibrations -- play a fundamental role in solid state physics and materials science.
They affect a wide variety of material properties and phenomena relevant for electronics, thermal transport, and optics \cite{Maldovan2013Nov,Li2012Jul} and are pivotal for quantum effects such as superconductivity.
Of particular interest are material systems in which phonons carry additional degrees of freedom, such as chirality \cite{Chen2018Dec,Zhu2018Feb,Liu2017Aug,Chen2019Mar}, polarization, or angular momentum \cite{Juraschek2019Jun,Cheng2020Aug}.
They are crucial for novel phonon-driven phenomena such as the phonon Hall effect \cite{Zhang2010Nov,Strohm2005Oct}, the phonon ac Stark \cite{Korenev2016Jan} and Edelstein effects \cite{Hamada2018Oct}, or the phonon Zeeman effect \cite{Juraschek2017Jun} and can even drive electronic phase transitions \cite{Forst2015Feb,Rini2007Sep,Nova2017Feb} and topological states \cite{Jotzu2014Nov,Nova2017Feb}.
A control over phonons and their degrees of freedom is thus a key goal on the pathway to potential phononic applications \cite{Maldovan2013Nov,Li2012Jul}, such as phonon-based quantum information devices \cite{Lee2011Dec}.
However, a microscopic control over phonons is very challenging due to their spin-less nature and the large inertia of the nuclei, which makes it hard to control them directly by electromagnetic fields.
In systems with strong electron-phonon interaction, however, a control over phonons can be achieved by manipulating the electrons.
In this regard, graphene is a prime candidate as it features strong electron-phonon interaction and allows for excellent external control over its electronic system.
In particular, the degenerate longitudinal (LO) and transverse optical (TO) phonons at the $\Gamma$~point of the first Brillouin zone are strongly coupled to the electronic system, which enables the tuning of their frequencies and lifetimes via an electrostatic gate \cite{Pisana2007}.
The electron-phonon coupling can further be significantly modified by a large external magnetic field, in which the electronic system condenses into discrete Landau levels, giving rise to so-called magneto-phonon resonances (MPRs) \cite{Neumann2015a,Yan2010,Kim2013May,Faugeras2011,Goler2012,Faugeras2009,Ando2007,Goerbig2007,Faugeras2018Jan,Sonntag2018May,Neumann2015Sep,Faugeras2012,Kossacki2012,Leszczynski2014,Kashuba2013Apr}.

\begin{figure*}
	\includegraphics[width=\textwidth]{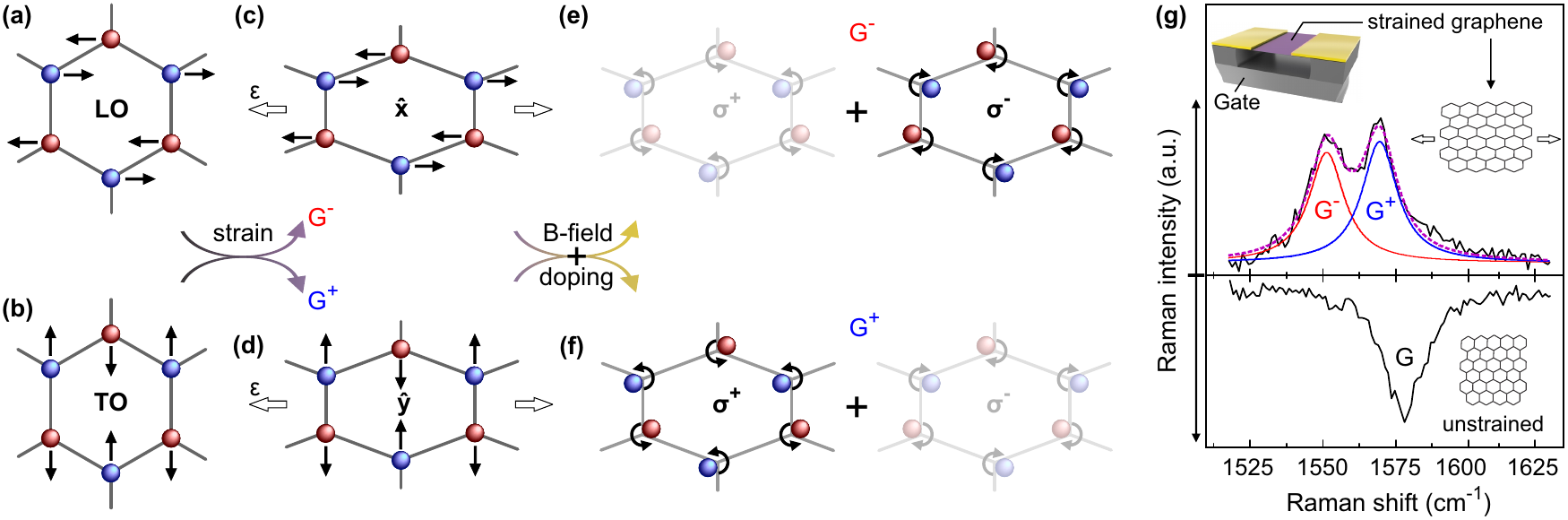}
	\caption{
	(a,b) Illustration of the vibrational pattern of the degenerate LO and TO phonon modes that give rise to the G~peak in pristine graphene.
	(c,d) The application of a uniaxial strain $\varepsilon$ leads to a splitting of the degenerate phonon modes into two branches G$^{\pm}$ that are linearly polarized along and perpendicular to the direction of strain.
	(e,f) The application of a static magnetic field $B$ in combination with a finite charge carrier density $n$ induces a net circular polarization of the phonon modes.
	The resulting vibrational patterns can be visualized as linear combinations of circular motions of the nuclei about their equilibrium positions with different weights for left- and right-handed movements, as indicated by the different levels of opaqueness.
	(g) Raman spectrum of graphene featuring the G~peak before (bottom) and after (top) the current annealing step.
	The purple line represents a fit to the two-subpeak structure with a sum of two Lorentzians (red and blue dotted lines).
	Upper left inset: Schematic illustration of graphene (purple) suspended from two gold contacts (yellow).
	The current annealing step induces uniaxial strain as illustrated by the two insets on the right.
	}
	\label{fig01}
\end{figure*}

Here we show that the control over the electron-phonon coupling in graphene allows to tune the movement of the nuclei from a linear oscillatory motion along or perpendicular to the bonds to a circular motion about their equilibrium positions.
We thus induce a circular polarization of the phonon modes, which yields control over the phonon angular momentum.
We achieve this purely via external static electric and magnetic fields in combination with uniaxial strain ($\sim 0.8 \%$), which lifts the degeneracy of the LO and TO phonon branches~\cite{Mohiuddin2009May,Mueller2017Nov} and facilitates the detection of their polarizations.
To monitor the phonon modes, we employ confocal Raman spectroscopy and directly extract the degree of circular phonon polarization from the measured peak positions and peak widths.
We quantitatively validate our findings with a microscopic model of the charge-carrier-dependent electron-phonon coupling.

We focus our attention on the LO and TO phonons of graphene at the $\Gamma$~point (see Figure~\ref{fig01}a,b).
In pristine graphene, these phonons are degenerate and give rise to the so-called G~peak in the Raman spectrum.
Under the application of uniaxial strain, this degeneracy is lifted and the phonon doublet splits into a softer, i.e., lower-frequency G$^-$ mode, which is linearly polarized in the direction of the strain, and a harder G$^+$ mode, which is linearly polarized perpendicular to the direction of the strain (see Figure~\ref{fig01}c,d) \cite{Mohiuddin2009May,Jung2019Mar}.
The strain-induced splitting allows us to track the Raman signal of each mode separately and, more importantly, to manipulate the vibrational pattern of the nuclei from a linear to a circular motion via tuning the electronic system and thus induce a circular phonon polarization.
We achieve the latter by applying an out-of-plane magnetic field $B$, which forces the electrons on cyclotron orbits and quantizes the electronic density of states into discrete Landau levels (LLs).
In a simple picture, this forces the nuclei into a circular motion due to their coupling to the electrons.
More precisely, the phonons couple to circularly polarized electronic transitions between the LLs.
At the charge neutrality point, the electron-hole symmetry implies that neither helicity, i.e., circular polarization, is preferred as electrons and holes rotate in opposite directions.
As a result, the phonons remain linearly polarized.
However, when tuning the charge carrier density $n$ with a gate voltage, we break this symmetry.
As a consequence, both phonon branches become increasingly circularly polarized with opposite helicity, i.e., the nuclei will increasingly adopt a circular motion about their equilibrium positions (see Figure~\ref{fig01}e,f).
The resulting phonon modes can be thought of as a linear superposition of the two helicities $\sigma^\pm$ with unequal weights, as indicated by the different levels of opaqueness in Figure~\ref{fig01}e,f.

Our experiments are based on a suspended graphene field effect device, see schematic in the inset of Figure~\ref{fig01}g.
It features a high charge carrier mobility and a low charge carrier density inhomogeneity, which are both crucial to observe magneto-phonon resonances.
A current-annealing step is employed to both induce uniaxial strain and to effectively clean the graphene sheet \cite{Bolotin2008}.
The cleaning results in a sharp peak in device resistance at the charge neutrality point with a charge carrier inhomogeneity of $n^*\approx5.5\times10^9$~cm$^{-2}$ (see  Figure~S1 in Supporting Information) and allows the observation of the quantum Hall effect for $B\approx0.1\,$T, indicating a charge carrier mobility of $\mu\approx100\:000~\mathrm{cm^2/(Vs)}$.
The presence of uniaxial strain after annealing is clearly visible in the splitting of the Raman G~peak into two separate peaks, see Figure~\ref{fig01}g.
To quantify the peak splitting and the induced strain, we fit the G~peak with a sum of two Lorentzians which share the same spectral width $\WG$.
The two subpeaks are centered around a mean frequency of $\bar{\omega}_\mathrm{G}\approx1566.4$~cm$^{-1}$ and split by $\Delta \wG\approx 14.8$~cm$^{-1}$.
From $\Delta \wG$ we estimate the amount of uniaxial strain in our sample as $\varepsilon\approx0.82$~\%, where we used the previously measured Gr\"uneisen parameters\cite{Mohiuddin2009May} $\partial \omega_\mathrm{G^+}/\partial \varepsilon \approx -18.6$~cm$^{-1}$/\% and $\partial \omega_\mathrm{G^-}/\partial \varepsilon \approx -36.4$~cm$^{-1}$/\%.
This significant amount of uniaxial strain induced by the current annealing step is in line with a recent report by Jung~\etal  \cite{Jung2019Mar}.
Note that we neglect biaxial strain, as it should be negligible in our device geometry and would result in the same shift for both peaks, which is irrelevant for the main experimental findings, i.e., the phonon polarization analysis.

\begin{figure*}
	\includegraphics[width=\textwidth]{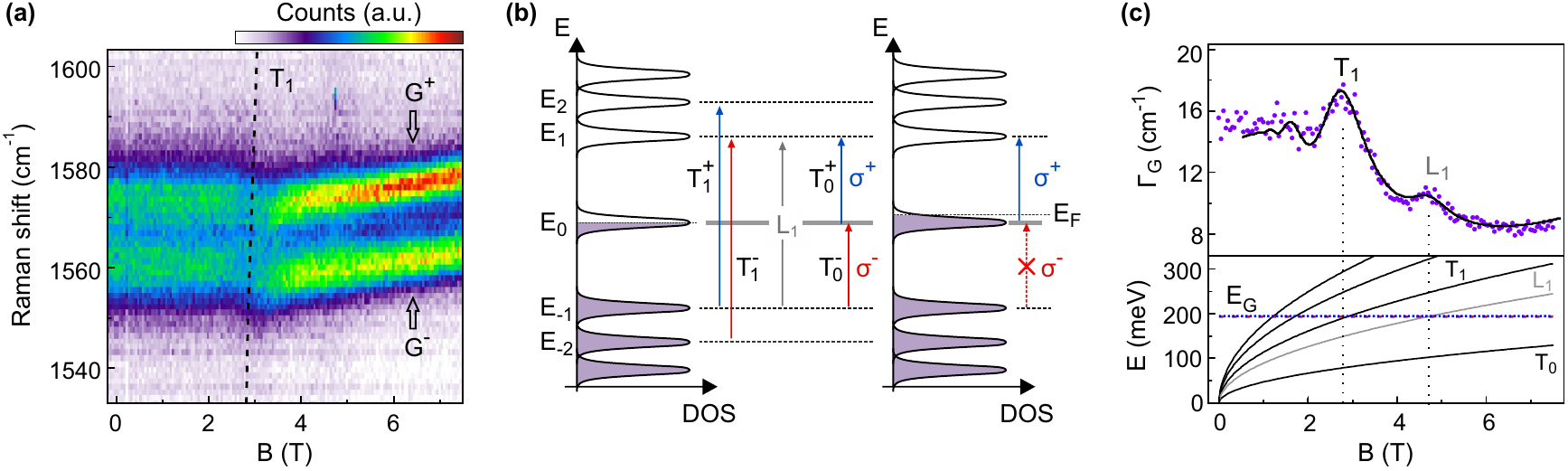}
	\caption{(a) Raman intensity around the strain-split G~peak of graphene as a function of $B$~field at $n\approx 0$ (highlighting the G$^-$ and G$^+$ peak).
	(b)~Illustration of the electronic transitions of the dominating MPRs and the density of states.
	Transitions with $\Delta j=(-)1$ only couple to right (left) circularly polarized phonons $\sigma^{+(-)}$, color coded as blue (red).
    (c) Lower panel: Interband LL transition energies $T_n$ (solid lines) and $L_1$ (grey, dotted lines) under the assumption of $v_\mathrm{\mathrm{T}_1}\approx1.35\times10^6$~m/s and $v_\mathrm{\mathrm{L}_1}\approx1.23\times10^6$~m/s.
    The closely spaced dashed red and dotted blue lines indicate the energies of the G$^-$ and G$^+$ mode phonons at $B=0$~T, respectively.
    The energy of the most relevant T$_1$ resonance is also shown in (a) as dashed line.
    Upper panel: $B$~field-dependent width $\WG$ of the two subpeaks G$^-$  and G$^+$.
    The black line represents a fit based on Dyson's equation.}
	\label{fig02}
\end{figure*}

We now turn to the control over the linearly polarized G~mode phonons via the manipulation of the electronic system.
To this end, we first focus on magneto-phonon resonances, i.e., the coupling of the phonons to the circularly polarized LL transitions, by measuring the $B$~field dependence of the linearly polarized modes at the charge neutrality point ($\EF=0$).
The measured Raman spectra for magnetic fields ranging from $0$ to $7.5$~T (and taken at $\approx 4.2$~K) are depicted in Figure~\ref{fig02}a.
Both the strain-induced splitting of the G~peak into the G$^+$ and G$^-$ peaks and the resonant coupling to LL transitions at $B\approx3$~T are clearly visible.
The magneto-phonon resonance at $B\approx3$~T is the so-called T$_1$~resonance \cite{Neumann2015a,Sonntag2018May,Ando2007,Goerbig2007}.
It occurs when the phonon energy, $E_\mathrm{G}=\hbar \bar{\omega}_\mathrm{G}(B=0,n=0)$, matches the energy difference between two LLs with energies $E_j = \pm v_{\mathrm{F}} \sqrt{2 e \hbar B \cdot j}$, where $v_\mathrm{F}$ is the  Fermi velocity and $j$ is the LL index.
The most prominent $\mathrm{T}_j$-MPRs involve LL transitions $-j\to j+1$ ($\mathrm{T}_j^{+}$) and $-(j+1)\to j$ ($\mathrm{T}_j^{-}$) with an orbital angular momentum of $\pm 1\,\hbar$ \cite{Ando2007,Goerbig2007}.
Importantly, angular momentum conservation implies that these transitions selectively couple to the respective circular phonon modes $\sigma^\pm$, as the latter also carry angular momentum $\pm1 \hbar$, due to the circular motion of the nuclei.
The corresponding transition energies $T_j$ and the resonance condition are then given by
\begin{equation}
T_j=E_{j+1}-E_{-j}=v_{\mathrm{T}_j} \sqrt{2 e \hbar B}(\sqrt{j+1}+\sqrt{j})=E_\mathrm{G}.
\label{eqStrainedMPR:Tj}
\end{equation}
The lower panel in Figure~\ref{fig02}c visualizes this resonance condition.
Note that we replaced $v_{\mathrm{F}}$ by the effective Fermi velocity $v_{\mathrm{T}_j}$, which can generally be a function of $n$, $B$, and the LL index $j$ due to many-body and excitonic effects \cite{Sonntag2018May,Faugeras2015,Chizhova2015Sep}.
To highlight the coupling of the two phonon modes to the LL transitions, we extract their frequency shifts $\Delta\omega_\mathrm{G^\pm}=\omega_\mathrm{G^\pm}-\omega_0^\pm$ relative to their respective frequencies at $B=0$~T as well as their shared width $\WG=\Gamma_\mathrm{G^+}=\Gamma_\mathrm{G^-}$  by fitting the sum of two Lorentzians.
Most noticeably, the coupling leads to a decrease of the phonon lifetime at the resonance due to the excitation of electron-hole pairs, which results in the increased width $\WG$ of both modes at the T$_1$ resonance, see upper panel in Figure~\ref{fig02}c.
We also observe a minor contribution of the L$_1$ ($-1 \to 1$) transition at $B\approx4.7$~T, which is usually forbidden due to the conservation of angular momentum, but can occur due to higher order processes \cite{Ando2007}.
The magnetic field $B_\mathrm{T_1}=2.73$~T, at which the T$_1$-MPRs occurs, corresponds to an effective Fermi velocity of $v_{\mathrm{T}_1}\approx1.35\times10^6$~m/s, as seen from Equation~\ref{eqStrainedMPR:Tj}.
Although strain has been predicted to change $v_\mathrm{F}$ \cite{Choi2010Feb,deJuan2012May}, we find that this value of $v_\mathrm{F}$ is in excellent agreement to previous results on the Fermi velocity renormalization in the presence of LLs in unstrained graphene \cite{Sonntag2018May}.

\begin{figure*}
	\includegraphics{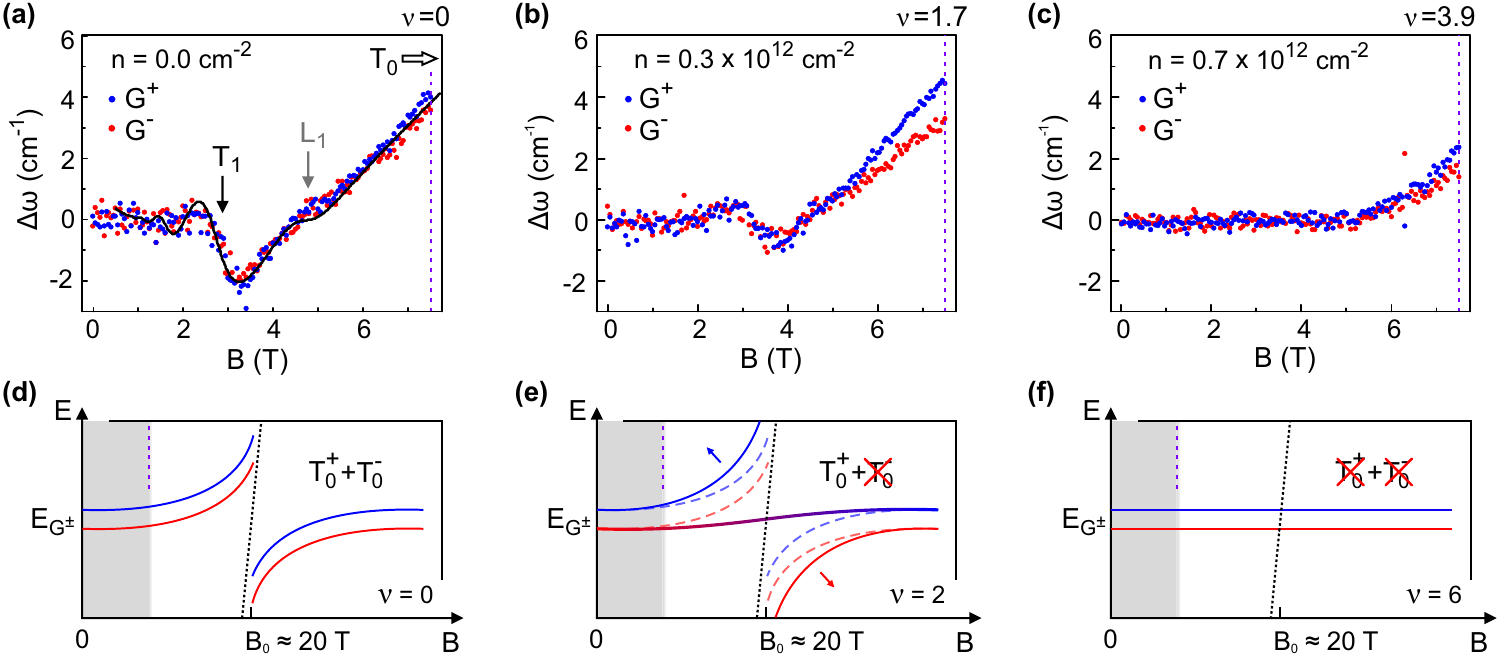}
	\caption{(a) Change in phonon frequency $\Delta\omega$ as a function of $B$ field for both the G$^-$ (red) and the G$^+$ mode (blue) at $n\approx0$.
	The absolute difference between the two modes is $14.8$~cm$^{-1}$.
	The black line represents a fit based on Dyson's equation.
	(b) and (c) show the same measurement as (a), but at intermediate and high $n$, respectively.
	(d-f) Illustration of the effect of the filling factor on the T$_0$-MPR.
	The colored lines are $E_\mathrm{G}^\pm = \hbar \wG^\pm$ and the dashed black line is $T_0$. At $\nu=2$ (e) the $\mathrm{T}_0^-$ transition is completely blocked while the $\mathrm{T}_0^+$ increases in strength.
	For comparison, the $\nu=0$ case shown in (d) is reproduced in (e) as colored dotted lines.
	(f) At $\nu=6$ all relevant LLs are completely filled and no transitions are possible.
	Note that $\nu$ is assumed to be constant for all $B$ for clarity.
	The area shaded in grey illustrates the measurement range shown in panels (a-c).
 }
	\label{fig03}
\end{figure*}

As shown in Figure~\ref{fig03}a, we observe identical magneto-phonon resonances of the two cross-linearly polarized G$^-$ and G$^+$ phonon modes at $n=0$.
Note that the splitting of $\Delta\wG=14.8$~cm$^{-1}$ is larger than the axis range in Figures~\ref{fig03}a-c.
This indicates that the strain-induced splitting dominates over the individual phonon frequency shift from the MPR.
This can be intuitively understood since at zero doping, the electron-hole symmetry is not broken and the T$_1^{\pm}$ LL~transitions remain degenerate and couple with equal strength to the phonons \cite{Kashuba2013Apr}.
As a result, there is no net circular phonon polarization at $n=0$.
In this regime, we can describe the MPRs by calculating the phonon self-energy and solving Dyson's equation \cite{Ando2007,Goerbig2007,Neumann2015a} (see Supporting Information for a more detailed discussion).
A combined fit of the solution of Dyson's equation to both $\omega_\mathrm{G^-}$ and $\WG$ is shown as black lines in Figure~\ref{fig02}d and Figure~\ref{fig03}a.
Evidently, the fit describes $\omega_\mathrm{G^+}$ just as well.

In the following, we show that a net circular phonon polarization can be induced by electrostatic gating, i.e., by tuning the system with a finite charge carrier density $n$.
After briefly discussing the influence of $n$ on $\mathrm{T}_1$, we will particularly focus on the gate-tunable off-resonant coupling of the phonons to the $\mathrm{T}_0^{\pm}$ transition, which occurs at large magnetic fields of around $B\approx 20$~T.
Here, the finite $n$ breaks the symmetry between $\mathrm{T}_0^{\pm}$, which results in a circular phonon dichroism~\cite{Kossacki2012,Leszczynski2014,Kashuba2013Apr} and thereby a net circular phonon polarization.

\begin{figure}
	\includegraphics{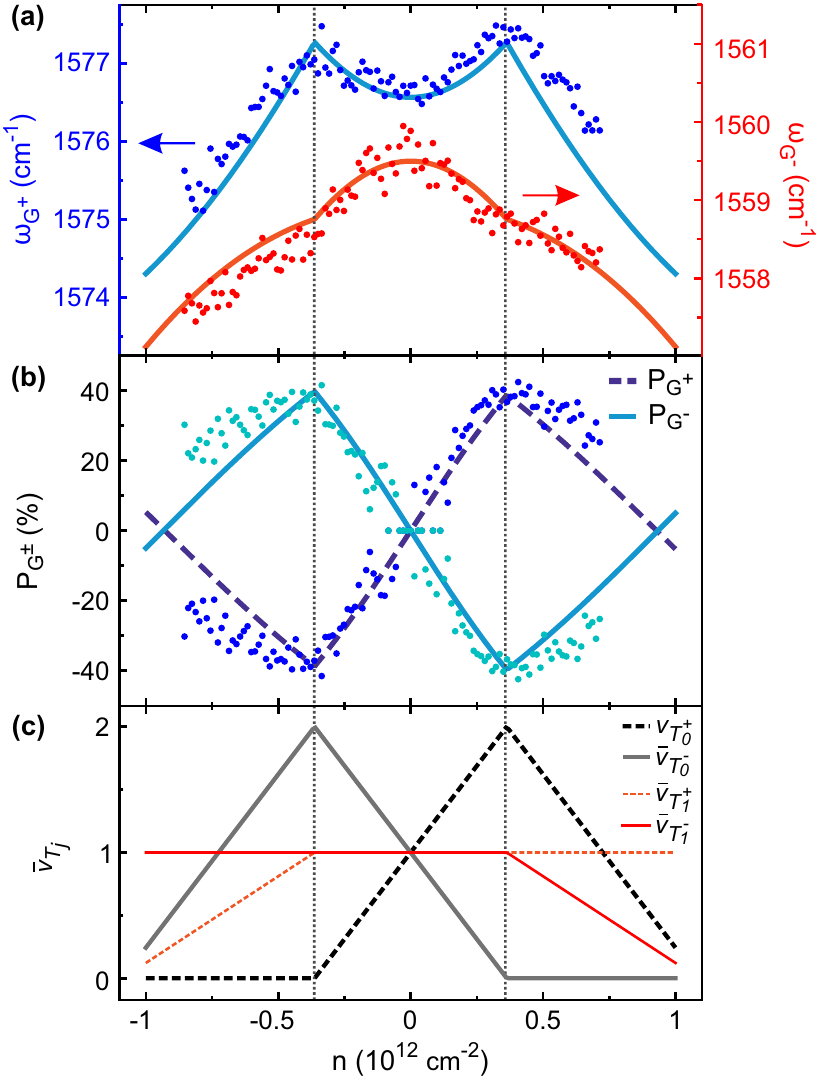}
	\caption{(a) Measured frequencies of the G$^+$ (blue dots, left axis) and the G$^-$ mode (red dots, right axis) as a function of $n$ (at $B=7.5$~T).
	The grey dotted lines mark a filling factor of $|\nu|=2$.
	The lines show $\omega_\mathrm{G^\pm}$ as calculated via the six-level model presented in the Supporting Information.
	(b)  Circular phonon polarization $P_\mathrm{G^+}$ (blue) and $P_\mathrm{G^-}$ (cyan) as calculated from the experimentally measured Raman peaks via Equation~\ref{eq:pol_model}.
	The lines correspond to the polarization as calculated via the six-level model presented in the Supporting Information.
	(c) Effective partial filling factor of the LL transitions $\bar{\nu}_{\mathrm{T}_j^\pm}$ as a function of $n$ at $B=7.5$~T.
	}
	\label{fig04}
\end{figure}

Figures~\ref{fig03}b,c show that the T$_1$-MPR gradually vanishes for higher charge carrier densities for both G$^-$ and G$^+$.
This is a consequence of the Pauli blocking of the respective transition (see Figure 2b).
An electron density of $n\approx0.7\times10^{12}$~cm$^{-2}$ corresponds to a filling factor $\nu = nh/(eB) \approx 10$ at $B\approx 3$~T, which leaves the second LL completely filled and blocks both the $-2\to 1$ and the $-1 \to 2$ transitions in Figure~\ref{fig03}c.
Notably, we observe a significant increase in $\Delta\wG$ at higher magnetic fields even at high~$|n|$, which we attribute to the tail of the T$_0$ resonance at $B\approx 20$~T.
Note that the T$_0$ resonance becomes unblocked as the effective filling factor decreases with $B$, see also purple dotted lines dotted lines in Figure~\ref{fig03}d-f.
Most strikingly, we observe that the evolution of the two phonon modes $\Delta\omega_\mathrm{G^-}$ and $\Delta\omega_\mathrm{G^+}$ split for large magnetic fields.
This is most pronounced at an intermediate electron density of $n\approx0.3\times10^{12}$~cm$^{-2}$ (see Figure~\ref{fig03}b).
The splitting can be attributed to the charge carrier-dependent change of the coupling strengths of the $\mathrm{T}_0^{\pm}$ transitions, which leads to a net circular phonon polarization.
For increasing $n$, the coupling to the $\mathrm{T}_0^{+}$ transition increases due to the increasing number of states partaking (see Fermi level shift for electron doping in Figure~\ref{fig02}b).
This leads to an enhanced anti-crossing for the $\sigma^{+}$ phonon branch, which is illustrated for a constant filling factor of $\nu= 2$ in Figure~\ref{fig03}d-f.
Compared to the situation for filling factor $\nu=0$, the enhanced anti-crossing of $\sigma^{+}$ with $\mathrm{T}_0^{+}$ leads to an increase in $\sigma^{+}$ frequency for $B$~fields below the resonance and consequently to an increased $\sigma^{+}$ circular polarization of the higher frequency peak G$^+$.
This holds up to a density where the second LL is completely filled and the coupling strength decreases again, resulting in a reduction of $\omega_\mathrm{G^+}$ for $\nu>2$.
In contrast, the $\mathrm{T}_0^{-}$ transition is suppressed by filling the zeroth LL (Figure~\ref{fig02}b), resulting in a weaker anti-crossing for $\sigma^{-}$ (see Figure~\ref{fig03}e), i.e., a monotonic decrease in $\omega_\mathrm{G^-}$ and an increased $\sigma^{-}$ polarization of G$^-$.
Indeed, as seen in Figure~\ref{fig04}a, both the G$^-$ and G$^+$ peaks show a qualitatively different dependence on $n$ at $B= 7.5$~T.
While $\omega_\mathrm{G^-}$ monotonically decreases in frequency with increasing $|n|$, $\omega_\mathrm{G^+}$ first increases for small $|n|$ before declining at larger densities.
As indicated by the grey dotted lines in Figure~\ref{fig04}a, the maxima in $\omega_\mathrm{G^+}$ are located precisely at the complete filling of the zeroth LL ($\nu=2$) at $n\approx0.3\times10^{12}$~cm$^{-2}$.

Most importantly, we can exploit the asymmetry between $\omega_\mathrm{G^\pm}(n)$ to directly calculate the degree of circular polarization of the phonon modes purely from the measured frequencies and widths.
We first note that purely circularly polarized phonons $\sigma^\pm$ can only couple to LL transitions of the same helicity due to angular momentum conservation.
The only coupling between phonons of different phonon helicities thus results from the uniaxial strain.
We can therefore describe the dynamics of the two phonon modes by an effective Hamiltonian
\begin{equation}
    H_{\mathrm{eff}}(n) = \hbar
    \begin{pmatrix}
       \tilde{\omega}_{\sigma^+}(n) & \eta \\
       \eta & \tilde{\omega}_{\sigma^-}(n)
    \end{pmatrix},
\end{equation}
written in the basis of the $\sigma^\pm$~modes, wherein $\tilde{\omega}_{\sigma^\pm}(n)=\omega_{\sigma^\pm}(n)-i \Gamma_{\sigma^\pm}(n)/2$ are the complex frequencies of the circular phonons $\sigma^{\pm}$, which include the frequency shifts from the interaction with LL~transitions.
The off-diagonal term $\eta$ describes the coupling between the two circular phonon modes due to strain.
We emphasize that this Hamiltonian is exact under the restrictions of angular momentum selection rules and does not depend on any microscopic model.
Instead, all quantities in $H_{\mathrm{eff}}$ can directly be obtained from the experimental data, as its eigenvalues $\hbar\tilde{\omega}_{\mathrm{G}^\pm}$ can directly be measured in form of the complex peak frequencies $\tilde{\omega}_{\mathrm{G}^\pm}=\omega_{\mathrm{G}^\pm}-i\Gamma_{\mathrm{G}^\pm}/2$.
Firstly, at $n=0$, we have $\tilde{\omega}_{\sigma^+}=\tilde{\omega}_{\sigma^-}$ due to the electron-hole symmetry, i.e., the splitting between the two measured phonon branches $\Delta\tilde{\omega}_\mathrm{G}=\tilde{\omega}_{\mathrm{G}^+}-\tilde{\omega}_{\mathrm{G}^-}$ is purely caused by the uniaxial strain and is equal to $2 \eta$.
Experimentally, we obtain $\eta = \Delta\tilde{\omega}_\mathrm{G}(n{=}0)/2  = 8.6\,\mathrm{cm}^{-1}$.
Secondly, $\tilde{\omega}_{\sigma^\pm}$ can be obtained from the eigenvalues $\hbar\tilde{\omega}_{\mathrm{G}^\pm}$ and $\eta$:
\begin{equation}
    \tilde{\omega}_{\sigma^\pm} = \frac{\tilde{\omega}_{\mathrm{G}^+}+\tilde{\omega}_{\mathrm{G}^-}}{2} \pm\frac{1}{2}\sqrt{\Delta\tilde{\omega}_\mathrm{G}^2 - 4\eta^2}.
\end{equation}
We can now determine the circular phonon polarization by calculating the eigenvectors $\mathbf{v}_{\mathrm{G}^{\pm}}$ of $H_{\mathrm{eff}}$ and project them on the circular basis vectors, i.e., on $\boldsymbol{\sigma}^+ = (1,0)^{\mathrm{T}}$ and $\boldsymbol{\sigma}^- = (0,1)^{\mathrm{T}}$:
\begin{equation}
    P_{\mathrm{G}^{\pm}} = \frac{|\boldsymbol{\sigma}^+ \cdot \mathbf{v}_{\mathrm{G}^{\pm}}|^2 - |\boldsymbol{\sigma}^- \cdot \mathbf{v}_{\mathrm{G}^{\pm}}|^2}
                                {|\boldsymbol{\sigma}^+ \cdot \mathbf{v}_{\mathrm{G}^{\pm}}|^2 + |\boldsymbol{\sigma}^- \cdot \mathbf{v}_{\mathrm{G}^{\pm}}|^2}.
\end{equation}
As the eigenvectors of $H_{\mathrm{eff}}$ can be calculated analytically, we are able to determine the circular phonon polarization purely from the measured phonon frequencies and widths:
\begin{equation}
    P_{\mathrm{G}^{\pm}} = \pm \frac{4\eta^2-|\Delta\tilde{\omega}_\mathrm{G}-\sqrt{(\Delta\tilde{\omega}_\mathrm{G})^2-4\eta^2}|^2}{4\eta^2+|\Delta\tilde{\omega}_\mathrm{G}-\sqrt{(\Delta\tilde{\omega}_\mathrm{G})^2-4\eta^2}|^2}.
    \label{eq:pol_model}
\end{equation}
The results of the conversion of the experimentally measured $\tilde{\omega}_{G^\pm}$ into circular polarization $P_{\mathrm{G}^{\pm}}$ are shown in Figure~\ref{fig04}b as blue and cyan dots, respectively.
For $n>0$, we find a near linear increase in $P_{\mathrm{G}^{+}}$ up to $\nu=2$, reaching circular polarization values of up to $\sim 40$~\%, before it decreases at larger $n$.
The behavior of $P_{\mathrm{G}^{-}}$ is exactly opposite and changes sign when tuning from electron ($n>0$) to hole ($n<0$) doping.
We emphasize that these $n$ dependent circular phonon polarizations are directly extracted from our measured Raman spectra and do not dependent on any knowledge of the nature of the electron-phonon coupling.

To gain a more detailed understanding of the microscopic origin of this polarization behavior, we will now look closer into the origin of the asymmetry in $\omega_{\sigma^\pm}(n)$.
As discussed above, the asymmetry can be traced back to the change in coupling strength of the circular phonon components $\sigma^\pm$ with the respectively polarized $\mathrm{T}_0^{\pm}$ LL~transitions.
In the following we show that the experimentally extracted phonon polarizations in Figure~\ref{fig04}b can be reproduced quantitatively by modeling the charge carrier-dependent electron-phonon coupling.
For this, we use an effective six-level model in which the electron-phonon system is described by six coupled quantum mechanical states, i.e., in which  we adopt a basis of the two circular phonon modes, the two $\mathrm{T}_0^{\pm}$, and the two $\mathrm{T}_1^{\pm}$ LL~transitions.
The details of this model are presented in the Supporting Information together with the full set of used parameters, which we extract from the fit shown in Figure~\ref{fig02}d and Figure~\ref{fig03}a at $n=0$.
Importantly, the coupling constants $g_{\mathrm{T}_j^\pm}$ between the phonons and the LL transitions depend on the filling of the LLs as $g_{\mathrm{T}_j^\pm}\propto \sqrt{\bar{\nu}_{\mathrm{T}_j^\pm}}$ and thus on $n$.
Here, the effective partial filling factor of the LL transition $\bar{\nu}_{\mathrm{T}_j^\pm}$, which is shown in Figure~\ref{fig04}c, is given by:
\begin{equation}
\begin{aligned}
\bar{\nu}_{\mathrm{T}_j^+}&= (1 + \delta_{j,0})(\pm \bar{\nu}_{\mp j} \mp \bar{\nu}_{\pm(j+1)}).
\end{aligned}
\label{eqStrainedMPR:FillingFactor}
\end{equation}
It is a function of the partial filling factor $\bar{\nu}_j$, which relates the filling factor $\nu$ to the fractional occupancy of the $j$th LL:
\begin{equation}
 	\bar{\nu}_j = \left.\frac{\nu+2-4j}{4}\right|_{\in[0,1]}.
 	\label{eq:FillingFactorPartial}
\end{equation}
In the context of this six-level model, we then calculate both $\omega_\mathrm{G^\pm}$ and the circular polarization $P_{\mathrm{G}^{\pm}}$ as a function of $n$ by diagonalizing the coupled electron-phonon system.

Evidently, we find good agreement between the six-level model and our experimental data, as shown in Figure~\ref{fig04}a~and~b.
We want to emphasize that the relevant parameters of the six-level model are extracted from the independent $B$~field measurement at $n=0$ (Figure 2d and Figure 3a), which makes the six-level model predictive for the measurements at fixed $B=7.5$~T and varying $n$.
Crucially, we observe the initial increase in $\omega_\mathrm{G^+}$ and $P_{\mathrm{G}^{+}}$ with $|n|$ and the kinks at $|\nu|=2$.
As discussed previously, the coupling of the circular $\sigma^{+(-)}$ phonon to the $\mathrm{T}_0^{+(-)}$-transition increases (decreases) with electron doping $n>0$ as more (less) electronic states become available,
compare the sketch in Figure~\ref{fig02}b and the corresponding effective LL transition filling factors $\bar{\nu}_\mathrm{\mathrm{T}_j}$ in Figure~\ref{fig04}c and their connection to the coupling constants $g_{\mathrm{T}_j^\pm}\propto \sqrt{\bar{\nu}_{\mathrm{T}_j^\pm}}$.
This leads to an increased $\sigma^{+}$~phonon frequency for any $B$ below the resonant magnetic field due to a stronger anti-crossing with the $\mathrm{T}_0^{+}$-transition, as illustrated in Figure~\ref{fig03}e.
Because the G$^+$~mode is defined as the higher frequency mode, its $\sigma^{+}$~polarization will consequently increase.
For $\nu>2$ the effective filling factor $\bar{\nu}_{\mathrm{T}_0^+}$ (see Figure~\ref{fig04}c) decreases and with it the coupling strength $g_{\mathrm{T}_0^+}$, which results in a weakening of the anti-crossing and thus in a decreased $\omega_\mathrm{G^+}$ and decreased circular phonon polarization.
Since the G$^-$~mode will always be predominantly polarized as the circular phonon with lower frequency, i.e., the circular phonon with the lower LL~transition coupling strength, it shows a monotonic decrease in $\omega_\mathrm{G^-}$ as a function of $n$ and a predominant $\sigma^-$ polarization.
It is important to emphasize that the $\mathrm{T}_0^{+(-)}$-transitions swap their roles when going from electron ($n>0$) to hole doping ($n<0$), i.e., for hole doping, the G$^+$~mode will be $\sigma^-$~polarized.
This results in the same polarization behavior, albeit with opposite sign.
Consequently, we are uniquely able to tune the circular polarization of the $\mathrm{G}^\pm$~modes, reaching polarization degrees of up to $\sim 40$~\%, purely by static electromagnetic fields due to the interplay of the strain-induced splitting of the phonon modes and the doping-induced asymmetry in electron-phonon coupling strength.
We are thus able to control the angular momentum of the phonons, which is directly related to the degree of polarization $P_\mathrm{G^\pm}$.

In conclusion, we explored the coupling of strain-split linearly polarized phonons to circularly polarized Landau level transitions in graphene.
We have shown that by electrostatic doping, we can selectively couple the phonons to right- or left-handed polarized Landau level transitions, which imprints their circular polarization onto the phonon modes.
This allows us to control the phonon polarization from a purely linear to a circular polarization of up to $40$~\%, purely by electrostatic fields.
This finding is verified by explicitly considering the charge carrier dependent electron-phonon coupling within an effective six-level model.
We expect that this novel control of the phonon polarization and of the phonon angular momentum by static electromagnetic fields will be of great interest for phononic applications \cite{Liu2017Aug,Zhu2018Feb} and will contribute to the rising field of research striving to manipulate phonons and their novel degree of freedoms~\cite{Chen2018Dec,Zhu2018Feb,Liu2017Aug,Juraschek2019Jun,Cheng2020Aug,Nova2017Feb,Chen2019Mar}.
Moreover, our observation of MPRs in strained graphene shows that MPRs can be a viable tool in the future to investigate strain gradient-induced pseudo-magnetic fields \cite{Verbiest2015}.

\begin{acknowledgments}
\textbf{Acknowledgments:}
The authors thank S. Staacks for help on the figures.
This project has received funding from the European Union’s Horizon 2020 research and innovation programme under grant agreement No 881603 (Graphene Flagship), the Deutsche Forschungsgemeinschaft (DFG, German Research Foundation) under Germany's Excellence Strategy - Cluster of Excellence Matter and Light for Quantum Computing (ML4Q) EXC 2004/1 - 390534769, through DFG (STA 1146/12-1)
and by the Helmholtz Nanoelectronic Facility (HNF)~\cite{Albrecht2017} at the Forschungszentrum J\"ulich.

\end{acknowledgments}


%%%%%%%%%% Merge with supplemental materials %%%%%%%%%%

\widetext
\pagebreak
\begin{center}
	\textbf{\large Supporting Information: Electrical Control over Phonon Polarization in Strained Graphene}
\end{center}
%%%%%%%%%% Merge with supplemental materials %%%%%%%%%%
%%%%%%%%%% Prefix a "S" to all equations, figures, tables and reset the counter %%%%%%%%%%
\setcounter{equation}{0}
\setcounter{figure}{0}
\setcounter{table}{0}
\setcounter{page}{1}
\makeatletter
\renewcommand{\theequation}{S\arabic{equation}}
\renewcommand{\thefigure}{S\arabic{figure}}
\renewcommand{\thesection}{S\arabic{section}}
\renewcommand{\thetable}{S\arabic{table}}

\section{Methods}
\textbf{Sample fabrication:}
The suspendended graphene sample is fabricated from an exfoliated graphene flake on Si/SiO$_2$.
The contacts made from Cr/Au  are fabricated via electron beam-lithography and a subsequent lift-off step.
The flake is suspended by etching away $\sim 170$~nm of SiO$_2$ with hydrofluoric acid.
Finally, a critical point drying-step is used to prevent the collapse of the suspended graphene flakes.
The graphene flake investigated here has a width of $2\,\mum$ and the distance between the contacts is $1.5\,\mum.$

\textbf{Raman measurements:}
The magneto-Raman measurements are performed in a confocal, low-temperature ($T=4.2$~K) micro-Raman setup, which is equipped with a superconducting magnet as well as electrical feedthroughs to conduct combined optical and transport experiments.
The laser ($\lambda=532\,\nm$) has a power of 0.5~mW and is focused to a spot size of $\sim500\,\nm$.
All Raman spectra are taken in the center of the graphene flake.

\section{Phonon self-energy for linear phonon polarization}

The evolution of the linearly polarized G~mode phonon frequency and width with magnetic field and doping can be obtained by solving Dyson's equation
\begin{equation}
    (\hbar\tilde{\omega})^2 - (\hbar\tilde{\omega}_0)^2 - 2\hbar\tilde{\omega}_0 \Pi(\tilde{\omega}) = 0
\end{equation}
for the complex phonon frequency $\tilde{\omega} = \wG - i \WG/2$.
Here, $\tilde{\omega}_0 = \omega_0 - i\gamma_{\mathrm{ph}}/2$, where $\omega_0$ is the phonon frequency in the absence of a magnetic field, at zero doping, and in the adiabatic approximation
and $\gamma_{\mathrm{ph}}$ accounts for the non-electronic contribution to the phonon decay width.
$\Pi(\tilde{\omega})$ denotes the phonon self-energy for linear phonon polarization and is the same for both $x$- and $y$-polarization.
It can be derived from the microscopic coupling of the Landau levels to the phonon within a tight binding-based model for the electron-phonon interaction~\cite{Ando2007,Goerbig2007,Neumann2015a} and can be written as
\begin{equation}
    \Pi(\tilde{\omega}) = \lambda\frac{T_0^2}{4} \sum_{s=\pm}\sum_{j=0}^{\infty} \left[ \bar{\nu}_{T^s_j}\frac{2\tilde{T}_j}{(\hbar\tilde{\omega})^2 - \tilde{T}_j^2} + \frac{2}{\tilde{T}_j}\right].
    \label{eq:SelfEnergyMPR}
\end{equation}
Here, $\tilde{T}_j = T_j - i\hbar\gamma_{\mathrm{T}_j}/2$ is the complex T$_j$ transition energy that includes a finite decay width $\gamma_{\mathrm{T}_j}$.
The partial filling factors $\bar{\nu}_{\mathrm{T}^s_j}$ have been defined in the main text, while $\lambda$ is a dimensionless effective electron-phonon coupling constant.
Finally, the experimentally observable, but in a first approximation symmetry-forbidden, L$_j$ ($-j \to +j$) transitions can be included phenomenologically via the replacement
\begin{equation}
    \Pi(\tilde{\omega}) \to \Pi(\tilde{\omega}) + \lambda_\mathrm{L} T_0^2\sum_{j=0}^{\infty} \left[ \bar{\nu}_{L_j}\frac{2\tilde{L}_j}{(\hbar\tilde{\omega})^2 - \tilde{L}_j^2} + \frac{2}{\tilde{L}_j}\right],
\end{equation}

with the respective partial filling factors being given by $\bar{\nu}_{\mathrm{L}_j} = (\bar{\nu}_{-j} - \bar{\nu}_{+j})$ and the complex transition energies being given by
$\tilde{L}_j = E_{+j} - E_{-j} -i\hbar\gamma_{\mathrm{L}_j}/2$.

\section{Details of the six-level model}
In the following, we introduce and discuss an effective six-level model to quantitatively describe the qualitatively different behaviour of $\omega_\mathrm{G^\pm}$.
Within the model, the coupled electron-phonon system is described by six coupled quantum mechanical states.
They represent the two phonon modes, the two $\mathrm{T}_0$-, and the two $\mathrm{T_1}$-LL excitations.
The phonons couple to the LL~transitions with coupling strengths $g_{T_j^{\pm}}$.
We choose to represent the phonon modes in the circular basis to easily include the selection rules resulting from the conservation of angular momentum, i.e., the $\sigma^\pm$ phonon only couples to the $\mathrm{T}_j^\pm$ ($-j \to j+1$) transitions.
In the basis $(\sigma^+,\sigma^-,\mathrm{T}_0^+,\mathrm{T}_0^-,\mathrm{T}_1^+,\mathrm{T}_1^-)$ the Hamiltonian is given by:
\begin{equation}
H= \begin{pmatrix}
	\hbar\omega_\mathrm{0} & \hbar\eta & g_\mathrm{T_0^+} & 0 &  g_\mathrm{T_1^+} & 0 \\
	\hbar\eta & \hbar\omega_\mathrm{0} & 0 & g_\mathrm{T_0^-} & 0 &  g_\mathrm{T_1^-} \\
	g_\mathrm{T_0^+} & 0 & T_0 & 0 & 0 & 0 \\
	0 & g_\mathrm{T_0^-} & 0 & T_0 & 0 & 0 \\
	g_\mathrm{T_1^+} & 0 & 0 & 0 & T_1 & 0 \\
	0 &  g_\mathrm{T_1^-} & 0 & 0 & 0 & T_1 \\
	\end{pmatrix}
	-i\frac{\hbar}{2}\Gamma,
	\label{eqStrainedMPR:SixLevel}
\end{equation}
where $\Gamma = \mathrm{diag} \left( \gamma_\mathrm{ph}, \gamma_\mathrm{ph}, \gamma_\mathrm{T_0}, \gamma_\mathrm{T_0}, \gamma_\mathrm{T_1}, \gamma_\mathrm{T_1}\right)$ is a diagonal matrix that describes the broadening of the phonons and electronic transitions.
$\hbar\omega_\mathrm{0}=E_\mathrm{G}$ and $T_j$ are the bare phonon and LL~transition energies, respectively.
$\eta$ is a parameter which describes the strain-induced splitting of the two modes.
The model

is based on the effective Hamiltonian
\begin{equation}
    \hat{H} = \sum_{s=\pm} \hbar{\omega_0} \hat{b}^{\dagger}_s\hat{b}_s
              + \hbar\eta \left( \hat{b}^{\dagger}_+\hat{b}_- + \hat{b}^{\dagger}_-\hat{b}_+ \right)
              + \sum_{\substack{s=\pm \\ j=0,1}} T_j \hat{d}^{\dagger}_{\mathrm{T}^s_j} \hat{d}_{\mathrm{T}^s_j}
              + \sum_{\substack{s=\pm \\ j=0,1}} g_{\mathrm{T}^s_j} \left( \hat{b}^{\dagger}_s \hat{d}_{\mathrm{T}^s_j} + \hat{d}^{\dagger}_{\mathrm{T}^s_j} \hat{b}_s \right),
    \label{eq:six-level-H}
\end{equation}
which is written in terms of the creation and annihilation operators for the circularly polarized phonon modes, $\hat{b}^{(\dagger)}_{\pm}$,
and for the $T_{j=0,1}$ LL transitions, $\hat{d}^{(\dagger)}_{\mathrm{T}^{\pm}_j}$.
The first term describes the uncoupled circularly polarized phonon modes, while the second term represents the strain-induced coupling between them.
The latter can be derived by considering a strain-induced phonon frequency splitting of $2\eta$ in a linear basis and then rotating the Hamiltonian in a linear basis into the circular basis.
The third term in Equation~\ref{eq:six-level-H} describes the uncoupled T$^{\pm}_j$ transitions, whereas the last term accounts for the electron-phonon coupling between phonons and the LL transitions, which is diagonal in the $\sigma^{\pm}$ basis.

The effective, doping-dependent coupling constants $g_{\mathrm{T}^{\pm}_j}$ can be derived by rotating the microscopic LL electron-phonon matrix elements for linear phonon polarization into the circular basis.
Alternatively, they can simply be read off by noting that the phonon self-energy for linear phonon polarization is the average of the phonon self-energies for circular polarization:
\begin{equation}
    \Pi(\omega) = \frac{1}{2} \left[ \Pi_{+}(\omega) + \Pi_{-}(\omega) \right]
                = \frac{1}{2} \sum_{s=\pm}\sum_{j=0}^{\infty} \left[ g_{\mathrm{T}_j^s}^2 \frac{2\tilde{T}_j}{(\hbar\omega)^2 - \tilde{T}_j^2} + \left.g_{\mathrm{T}_j^s}^2\right|_{n=0} \frac{2}{\tilde{T}_j}\right],
\end{equation}
from which we can identify
\begin{equation}
    g_{\mathrm{T}^{\pm}_j} =  T_0 \sqrt{\bar{\nu}_{\mathrm{T}^{\pm}_j} \lambda /2} \,= \vF\sqrt{\bar{\nu}_{\mathrm{T}_j^\pm}\lambda e\hbar B}.
\end{equation}

Here, the effective partial filling factors of the LL~transitions, $\bar{\nu}_{\mathrm{T}_j^\pm}$, are a function of the magnetic field and are given by:
\begin{equation}
\begin{aligned}
\bar{\nu}_{\mathrm{T}_j^+}&= (1 + \delta_{j,0})(\bar{\nu}_{-j} - \bar{\nu}_{j+1}), \\
\bar{\nu}_{\mathrm{T}_j^-}&= (1 + \delta_{j,0})(\bar{\nu}_{-(j+1)} - \bar{\nu}_{+j}).
\end{aligned}
\label{eqStrainedMPR:FillingFactor_Supp}
\end{equation}
The partial filling factor $\bar{\nu}_j$ relates the filling factor $\nu$ to the fractional occupancy of the $j$th LL:
\begin{equation}
 	\bar{\nu}_j = \left.\frac{\nu+2-4j}{4}\right|_{\in[0,1]},
 	\label{eq:FillingFactorPartial_Supp}
\end{equation}
which is defined as $\bar{\nu}_j=0$ if ${\nu+2-4j}/{4}<0$ and $\bar{\nu}_j=1$ if ${\nu+2-4j}/{4}>1$ and considers the spin and valley degeneracy of graphene.
We are now able to calculate the energies of the phonon modes under the influence of the coupling to the $\mathrm{T}_0^\pm$ and $\mathrm{T}_1^\pm$ transitions by numerically diagonalizing Equation~\ref{eqStrainedMPR:SixLevel}.
To this end, we use the experimental parameters extracted from the fit shown in Figure~2d and Figure~3a of the main manuscript at $n=0$, which results in effective Fermi velocities $v_\mathrm{\mathrm{T}_1}\approx1.35\times10^6$~m/s and $v_\mathrm{\mathrm{L}_1}\approx1.23\times10^6$~m/s.
For a more stable fit, we assumed the same $v_\mathrm{\mathrm{T}_j}=v_\mathrm{\mathrm{T}_1}$ for all $j$ expect for $j=0$, which we set to $v_\mathrm{\mathrm{T}_0}=v_\mathrm{\mathrm{L}_1}$, since $\mathrm{T}_0$ shares the first LL with $\mathrm{L}_0$.
The other extracted parameters are the electronic broadening of the $\mathrm{T}_j$-transitions $\gamma_{\mathrm{T}_j}\approx392$~cm$^{-1}$ and $\mathrm{L}_1$-transition $\gamma_{\mathrm{L}_1}\approx196$~cm$^{-1}$, a phonon broadening $\gamma_{\mathrm{ph}}\approx2.5$~cm$^{-1}$, an electron-phonon coupling constant $\lambda = 4.5 \times 10^{-3}$ \cite{Pisana2007,Froehlicher2015May}, and a coupling strength of the $\mathrm{L}_1$-transition $\lambda_{\mathrm{L}}\approx0.27\times10^{-3}$.
Further, within this model, we assume $v_\mathrm{T_0}=v_\mathrm{L_1}$, $\gamma_\mathrm{T_0}=\gamma_\mathrm{L_1}$ and use $\eta=8.6$~cm$^{-1}$, and $\omega_0 = 1566.4\,\cm$.

To calculate the net circular phonon polarization, we first obtain the weight of the $\sigma^{\pm}$ component in the $\mathrm{G}^{\pm}$~mode by projection on the respective component: $w(\sigma^{\pm},\mathrm{G}^{\pm}) = |\boldsymbol{\sigma}^{\pm} \cdot \mathbf{v}_{\mathrm{G}^{\pm}}|^2$.
We then define the circular polarization of the $\mathrm{G}^{\pm}$~modes as
\begin{equation}
    P_{\mathrm{G}^{\pm}} = \frac{w(\sigma^+,\mathrm{G}^{\pm}) - w(\sigma^-,\mathrm{G}^{\pm})}{w(\sigma^+,\mathrm{G}^{\pm}) + w(\sigma^-,\mathrm{G}^{\pm})}
                = 2 w(\sigma^+,\mathrm{G}^{\pm}) - 1.
    \label{eq:polar}
\end{equation}
Here, we made use of the fact that both weights need to add up to one: $w(\sigma^+,\mathrm{G}^{\pm}) + w(\sigma^-,\mathrm{G}^{\pm}) = 1$.
Note that in the interacting electron-phonon system, this identity does no longer hold exactly, as the phonon states can now contain an admixture of electronic excitations, but it is still a very good approximation in the off-resonant electron-phonon coupling regime considered here.

\section{Additional electrical transport and Raman measurements}

\begin{figure}[H]
    \centering
	\includegraphics{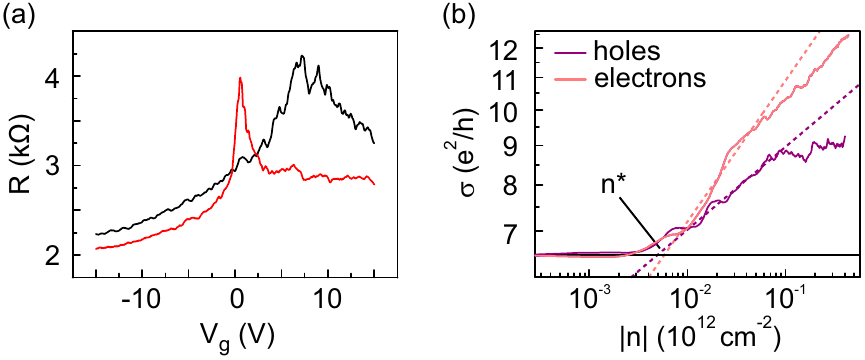}
	\caption{(a) The resistance as a function of $V_\mathrm{g}$ before (black) and after current annealing (red). (b) Double logarithmic plot of $\sigma$ after current annealing versus charge carrier density $|n|$ for determining the charge carrier inhomogeneity $n^*$.
	}
	\label{figS01}
\end{figure}

\begin{figure}[H]
    \centering
	\includegraphics{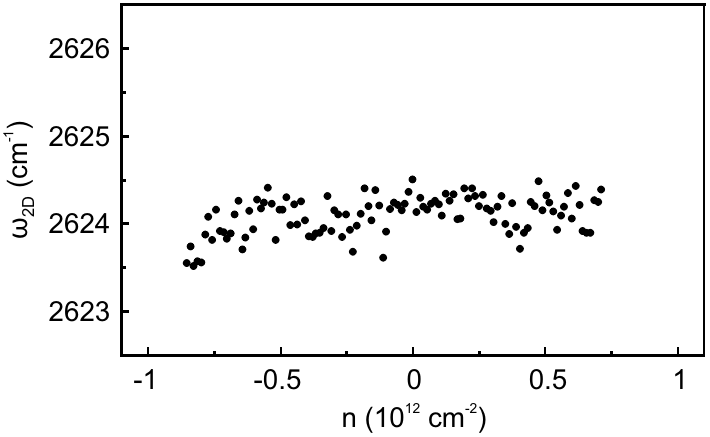}
	\caption{The 2D~peak position $\wD$ does not show significant variation as a function of $n$, indicating a negligible amount of gate-induced strain.
	}
	\label{figS02}
\end{figure}

\begin{figure}[H]
    \centering
	\includegraphics{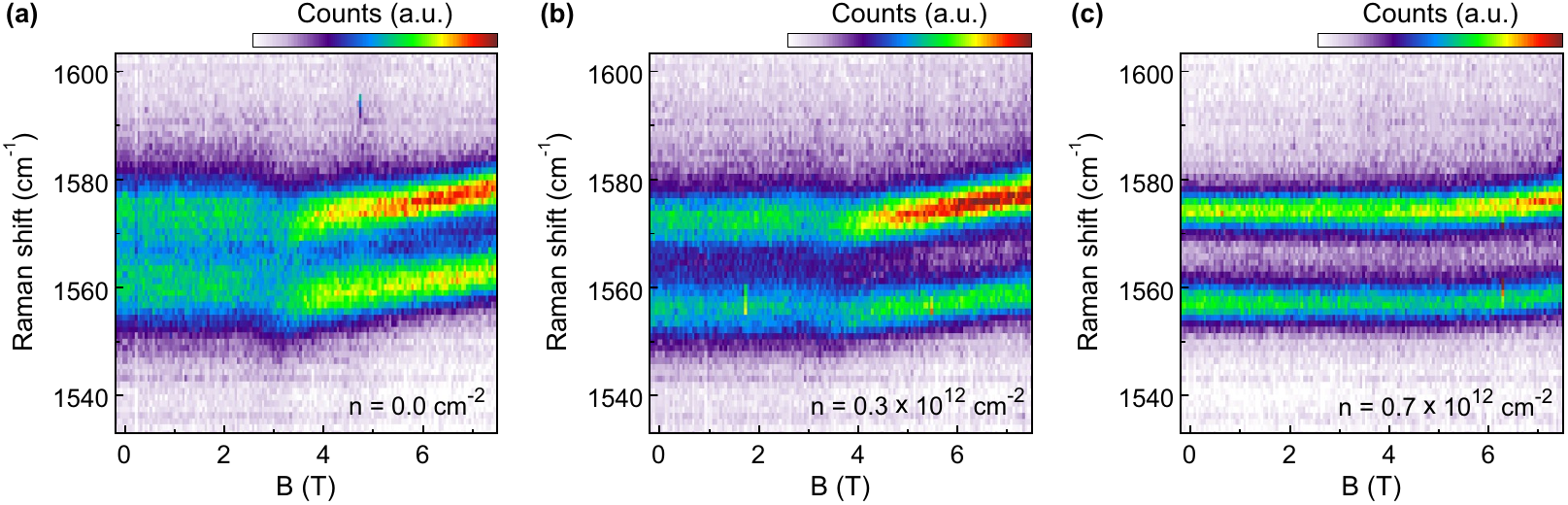}
	\caption{(a-c) Raman intensity around the strain-split G~peak of graphene as a function of $B$~field at $n\approx 0~\mathrm{cm}^{-2}$, $n=0.3\times10^{12}~\mathrm{cm}^{-2}$, and $n=0.7\times10^{12}~\mathrm{cm}^{-2}$, respectively. The measurement corresponds to the data shown in Figure~3a-c of the main manuscript. (a) is reprinted from Figure~2a for comparison.
	}
	\label{figS03}
\end{figure}

\end{document}